Territoires intelligents durables et approche multi-agents :
Retour d'expériences de travaux menés à l'île de La Réunion

Rémy Courdier, Laboratoire d'Informatique et de Mathématiques, Université de La Réunion, France

**Résumé**

La Réunion est une île européenne située au cœur de l'océan indien, éloignée des ressources du territoire de la France métropolitaine. C'est une île riche en diversité qui doit faire face à de nombreux enjeux sociétaux. Ainsi, de nombreux acteurs travaillent ensemble afin d'accompagner et soutenir la transformation de son modèle sociétale vers un modèle qui repose sur la résilience territoriale et collective. Nous proposons dans cet article de faire un retour sur les études et expérimentations menées depuis plusieurs années à La Réunion par un groupe de recherche du domaine de la simulation multi-agents et de montrer par des exemples de quelle manière cette approche peut contribuer très concrètement aux processus de compréhension et de fabrication de modèles comportementaux et d'aménagement de l'espace d'un territoire. Nous commencerons par poser les principes de base de l'approche multi-agents, puis nous présenterons nos retours d'expériences issus des expérimentations sur des domaines diversifiés dans la prospective territoriale. L'article se termine en montrant la démarche mise en œuvre par l'équipe de recherche pour tisser le lien entre les acteurs de la transformation du territoire au travers de l'animation d'un cercle de réflexions.

# 1  Introduction : La Réunion, territoire d'expérimentations

L'île de La Réunion est une île volcanique de 2512 km2 située dans l'océan Indien occidental, dont l'altitude varie du niveau de la mer à 3070 m. Actuellement, plus de 80 % des 863 100 habitants vivent sur la frange côtière, où se concentrent la plupart des activités socio-économiques. La population est jeune (2,5 fois plus de moins de 20 ans que de personnes de 65 ans ou plus) et augmente de 0,4 % en moyenne par an depuis une dizaine d'année [1]. Cette population est constituée d'un large mélange de cultures. Depuis la création d'un parc national en 2007, 43 % de la surface de l'île est protégée par des réserves spécifiquement dédiées à la conservation de la biodiversité. La pression sur les espaces engendrés par le parc pose la question de la soutenabilité environnementale de l'île, avec notamment l'enjeu de contenir l'artificialisation des sols, déjà particulièrement élevée sur la bande littorale, ce qui provoque nécessairement de fortes tensions entre les parties prenantes sur la planification et la gestion de l'espace [2].

Ainsi, La Réunion concentre les principaux défis actuels de l'aménagement des territoires sur une région aux frontières bien délimitées par son isolement géographique, au travers d'un ensemble d'acteurs aisément identifiables dans chaque secteur d'activité. Par ailleurs, La Réunion accueille de nombreux organismes de recherche produisant de grandes quantités de données utilisables pour accompagner les décideurs sur les grands défis de l'île qui sont principalement : la maîtrise de l'étalement urbain, l'adaptation des infrastructures (notamment les routes et les dispositifs d'adduction d'eau), le développement des transports en commun, la gestion des besoins énergétiques, la protection des terres agricoles et la conservation de la biodiversité.

Sa diversité culturelle et géographique, son insularité et ses nombreux défis, confère à l'île de La Réunion un cadre tout particulièrement favorable comme territoire d'expérimentation en matière d'aménagement de territoires intelligents et notamment pour participer au débat des enjeux posés par les technologies numériques associées.

# 2  La simulation multi-agents

La simulation consiste à réaliser une représentation informatique d'un système et de son comportement au cours du temps au moyen de modèles ; cette démarche s'avère particulièrement appropriée dès qu'un système n'est pas accessible directement à l'observation ou à la mesure. La simulation multi-agents offre un modèle informatique qui vise à représenter directement les entités, leurs comportements propres et leurs interactions. De cette manière, il devient possible d'analyser un phénomène comme le résultat d'interactions entre des entités autonomes, dénommées agents. La simulation multi-agents permet de construire de véritables micro-mondes artificiels dont on peut contrôler tous les paramètres (quantitatifs ou qualitatifs) et cela à tous les niveaux du système, qu'il s'agisse de l'entité, du groupe, de la société ou au niveau des effets externes sur l'environnement. L'un des intérêts majeurs de la simulation multi-agents tient à ce que l'on puisse expérimenter directement sur des modèles réduits de sociétés les théories ou alternatives que l'on souhaite étudier [3].

L'objectif de notre groupe de recherche est de proposer des théories, méthodes et outils facilitant les interactions entre les informaticiens et les thématiciens qui apportent leur expertise à la construction de modèles de simulation. Il s'agit de faire en sorte que de tels modèles puissent fournir des pistes de réflexions aux décideurs confrontés aux choix de demain afin d'aider à mettre en place une organisation des territoires qui soit la plus cohérente possible.

Pour cela nous avons travaillé sur un certain nombre de propositions relatives à la conception de plateformes de simulation et de modélisation de systèmes multi-agents adaptés à la gestion territoriale. Nous avons retenu ci-dessous quatre contributions principales :

(1) L'un de nos premiers objectifs a été d'intégrer la complexité de la prise de décision propre à l'Homme dans de tels outils, à ce sujet nous avons proposé l'adaptation du modèle de pyramide de besoins proposée par le psycho-sociologue Abraham Maslow afin de pouvoir représenter dans nos modèles de simulation une catégorisation et une hiérarchisation des besoins chez l'être humain (besoins physiologiques, de sécurité, sociaux, d'estime et d'auto-réalisation) [4].

(2) Nous avons constaté que dans ce type de contexte comprenant de nombreux interlocuteurs, il est important de faire en sorte que l'information utilisée dans les modèles soit comprise par tous de la même façon. Nous avons alors proposé une technique d'initialisation des modèles de simulation reposant sur l'utilisation de cartes reposant sur des codes couleur sémantiques qui font office de plans d'informations. Ces cartes s'avèrent être des vecteurs de transmission de l'information pour les différents thématiciens, entre eux, et vers les modélisateurs qui participent avec eux à la construction d'outils de simulation [5].

(3) Par ailleurs, nous avons travaillé sur la place qui est accordée dans les modèles de gestion territoriale à la notion d'émergence de phénomènes et souligné la nécessité, pour les concepteurs des modèles, de bien distinguer, d'une part, l'émergence de phénomènes et, d'autre part, la capacité à détecter, caractériser et manipuler ces mêmes phénomènes dans un outil de simulation [6].

(4) Nous avons également montré l'intérêt d'un modèle d'agent basé sur la théorie de l'action située pour la simulation d'entités situées dans l'espace. Alors que l'approche classique met l'accent sur la décision de l'acteur, l'action située est considérée comme un processus doté d'une épaisseur temporelle, émergeant spontanément des situations créées par les interactions locales entre l'acteur et son environnement. Ce modèle, qui prend en compte les dimensions temporelles et spatiales de l'action ainsi que ses caractéristiques contingentes, met en œuvre les concepts d'affordance (capacité des objets à déclencher des actions) et de stigmergie (auto-organisation médiée par les marques laissées par les individus dans leur environnement) [7].

## 3   Retour d'expérience sur les expérimentations

Compte tenu des défis d'aménagement du territoire présentés dans l'introduction, la mise en place d'outils d'aide à la décision s'avère ici être une nécessité afin de pouvoir étudier les alternatives possibles pour les avenirs de l'île. La Réunion s'est révélée un terrain d'expérimentation particulièrement riche pour illustrer les apports de la simulation agents relativement à de nombreuses problématiques de gestion territoriale tels que : la pression démographique sur l'espace [4], la mobilité urbaine, le besoin énergétique [8], la gestion des bio-déchets agricole et urbain [9], la gouvernance des ressources communes [10], l'aide à la mobilité urbaine de personnes en perte d'autonomie [11], tout en considérant des solutions numériques éco-responsables [12].

Le retour d'expérience relativement à ces travaux, met en évidence plusieurs constats :

- Des outils avancés d'analyse des interactions des agents participant aux simulations sont nécessaires pour comprendre l'apparition de phénomènes et pour contribuer à la validation de systèmes de simulation ;
- Les cartes utilisant des codes de couleur sémantiques sont des supports très pertinents pour construire des environnements multi-agents : elles sont faciles à manipuler par les informaticiens et les thématiciens ;

- Le plus important est moins la solution que le processus qui y conduit. L'objectif est donc d'améliorer l'apprentissage social et de générer des connaissances sociales pour une prise de décision éclairée ;
- Un développement participatif doit être au cœur de la démarche afin d'explorer des scénarios alternatifs pour gérer les conflits des acteurs territoriaux ;
- La constitution d'un système multi-agents d'aide à la gestion territoriale n'est pas une tâche simple. La construction d'une méthodologie nécessite une approche pluridisciplinaire sur des applications réelles.

## 4   Tisser le lien avec les acteurs de la recherche et de la société civile

Dans ce type de contexte comprenant de nombreux interlocuteurs, notre groupe de recherche a initié un cercle de réflexion [13] pour partager les expériences et mener des échanges ouverts ayant les objectifs suivants :

(1) Identifier les acteurs de la fabrication des territoires intelligents de l'île ;
(2) Dresser un inventaire des actions menées par les différents acteurs ;
(3) Identifier les synergies entre acteurs et les sujets qui peuvent être travaillés collectivement ;
(4) Partager les expériences sur des sujets tels que la sécurité, la robustesse, la résilience et les standards ;
(5) Encourager les entreprises à ouvrir leurs données et stimuler la création de services intersectoriels ;
(6) Stimuler la création de l'écosystème d'un territoire numérique dans une démarche de cohérence.

Ce cercle de réflexion a permis la création, avec les acteurs du territoire, d'un ensemble de vidéo pédagogiques regroupées dans la série « Ville intelligente » de la Collections « Sciences et Société » produite par l'Université Numériques de La Réunion (UNR) et diffusés sur les réseaux sociaux universitaires [14].

## 5   Références